\documentclass[dvips]{acta}
\usepackage{supertabular,lscape,epsfig}
\usepackage{amssymb}
\usepackage{amsmath}
\usepackage{float}
\mathchardef\mhyphen="2D
\usepackage{placeins}%\Floatbarier
\DeclareSymbolFont{ppa}{OT1}{ppl}{m}{it}
\DeclareMathSymbol{\vv}{\mathalpha}{ppa}{'166}

\SetPages{1}{19}

\SetVol{70}{2020}

\begin{document}
\newcommand\pvalue{\mathop{p\mhyphen {\rm value}}}
%Zwarte naglowki, jeden wiersz, appendix
\newcommand{\TabApp}[2]{\begin{center}\parbox[t]{#1}{\centerline{
  {\bf Appendix}}
  \vskip2mm
  \centerline{\small {\spaceskip 2pt plus 1pt minus 1pt T a b l e}
  \refstepcounter{table}\thetable}
  \vskip2mm
  \centerline{\footnotesize #2}}
  \vskip3mm
\end{center}}

%Zwarte naglowki, jeden wiersz
\newcommand{\TabCapp}[2]{\begin{center}\parbox[t]{#1}{\centerline{
  \small {\spaceskip 2pt plus 1pt minus 1pt T a b l e}
  \refstepcounter{table}\thetable}
  \vskip2mm
  \centerline{\footnotesize #2}}
  \vskip3mm
\end{center}}

%Zwarte naglowki, dwa wiersze
\newcommand{\TTabCap}[3]{\begin{center}\parbox[t]{#1}{\centerline{
  \small {\spaceskip 2pt plus 1pt minus 1pt T a b l e}
  \refstepcounter{table}\thetable}
  \vskip2mm
  \centerline{\footnotesize #2}
  \centerline{\footnotesize #3}}
  \vskip1mm
\end{center}}

%wewn±trz minipage
\newcommand{\MakeTableH}[4]{\begin{table}[H]\TabCap{#2}{#3}
  \begin{center} \TableFont \begin{tabular}{#1} #4 
  \end{tabular}\end{center}\end{table}}

%Zwarte naglowki, jeden wiersz, appendix
\newcommand{\MakeTableApp}[4]{\begin{table}[p]\TabApp{#2}{#3}
  \begin{center} \TableFont \begin{tabular}{#1} #4 
  \end{tabular}\end{center}\end{table}}

%Zwarte naglowki, jeden wiersz
\newcommand{\MakeTableSepp}[4]{\begin{table}[p]\TabCapp{#2}{#3}
  \begin{center} \TableFont \begin{tabular}{#1} #4 
  \end{tabular}\end{center}\end{table}}

%Zwarte naglowki, jeden wiersz
\newcommand{\MakeTableee}[4]{\begin{table}[htb]\TabCapp{#2}{#3}
  \begin{center} \TableFont \begin{tabular}{#1} #4
  \end{tabular}\end{center}\end{table}}

%Zwarte naglowki, dwa wiersze
\newcommand{\MakeTablee}[5]{\begin{table}[htb]\TTabCap{#2}{#3}{#4}
  \begin{center} \TableFont \begin{tabular}{#1} #5 
  \end{tabular}\end{center}\end{table}}

%Tabela w okre¶lonym miejscu
%\newcommand{\MakeTableH}[4]{\begin{table}[H]\TabCap{#2}{#3}
%  \begin{center} \TableFont \begin{tabular}{#1} #4 
%  \end{tabular}\end{center}\end{table}}

%Tabela w okre¶lonym miejscu, zwatre nag³ówki, jeden wiersz
\newcommand{\MakeTableHH}[4]{\begin{table}[H]\TabCapp{#2}{#3}
  \begin{center} \TableFont \begin{tabular}{#1} #4 
  \end{tabular}\end{center}\end{table}}

%wyrównanie w tabeli - wzglêdem kropki dziesiêtnej r@.l (a2_19/osb - ttt9)
%{\it Acta Astronomica Archive}
%\parskip=0pt \itemsep=1mm \setlength{\itemsep}{0.4mm}\setlength{\parindent}{-1em} \setlength{\itemindent}{-1em} - po \begin{itemize} - wszystko
%FWHM, PSF, S/N - proste, 
%MgII, H$\alpha$
%rms, rhs, sd - kursywa
%{\sc DAOPhot}
%{\sc Fnpeaks}
%{\sf files}
%Galactic wszystko (bulge, center, plane, disk, coordinates, latitudes...)
%Cepheids
%type~ Cepheids, Population~II Cepheids
%a.u. => au (od AcA 3/2018)
%Polish National Science Centre
\newfont{\bb}{ptmbi8t at 12pt}
\newfont{\bbb}{cmbxti10}
\newfont{\bbbb}{cmbxti10 at 9pt}
\newcommand{\uprule}{\rule{0pt}{2.5ex}}
\newcommand{\douprule}{\rule[-2ex]{0pt}{4.5ex}}
\newcommand{\dorule}{\rule[-2ex]{0pt}{2ex}}
\def\thefootnote{\fnsymbol{footnote}}
\begin{Titlepage}
\Title{Additional Galactic Cepheids from the OGLE Survey\footnote{Based on observations obtained with 
the 1.3-m Warsaw telescope at the Las Campanas Observatory of the 
Carnegie Institution for Science.}}
\vspace*{-3pt}
\Author{I.~~S~o~s~z~y~ñ~s~k~i$^1$,~~
A.~~U~d~a~l~s~k~i$^1$,~~
M.\,K.~~S~z~y~m~a~ñ~s~k~i$^1$,~~
P.~~P~i~e~t~r~u~k~o~w~i~c~z$^1$,\\
J.~~S~k~o~w~r~o~n$^1$,~~
D.\,M.~~S~k~o~w~r~o~n$^1$,~~
R.~~P~o~l~e~s~k~i$^1$,~~
S.~~K~o~z~³~o~w~s~k~i$^1$,\\
P.~~M~r~ó~z$^{1,2}$,~~
K.~~U~l~a~c~z~y~k$^{3,1}$,~~
K.~~R~y~b~i~c~k~i$^1$,~~
P.~~I~w~a~n~e~k$^1$,~~
M.~~W~r~o~n~a$^1$~~\\
and~~M.~~G~r~o~m~a~d~z~k~i$^1$
}
{$^1$Astronomical Observatory, University of Warsaw, Al.~Ujazdowskie~4, 00-478~Warszawa, Poland\\
$^2$Division of Physics, Mathematics, and Astronomy, California Institute of Technology, Pasadena, CA 91125, USA\\
$^3$Department of Physics, University of Warwick, Gibbet Hill Road, Coventry, CV4~7AL,~UK}
\Received{June 18, 2020}
\end{Titlepage}

\vspace*{-5pt}
\Abstract{We report on the results of a comprehensive search for
  Cepheid variable stars in the fields photometrically monitored by
  the OGLE Galaxy Variability Survey. We identify 742 Cepheids not
  included in the first release of the OGLE Collection of Galactic
  Cepheids and reclassify several dozen previously published
  variables. The upgraded collection comprises 1974 classical, 1625
  type~II, and 119 anomalous Cepheids located in the area of about
  3000 square degrees covering the disk and bulge of the Milky Way.

  We present the most interesting objects in our sample: an isolated
  group of long-period double-mode classical Cepheids pulsating in the
  fundamental mode and first overtone, the first known Galactic
  double-mode Cepheid with the second and third overtone excited,
  double-mode type~II Cepheids, candidates for single-mode
  first-overtone BL~Her stars, and Cepheids showing simultaneous
  eclipsing variability. We also discuss type~II and anomalous
  Cepheids that are potential members of globular clusters.}{Stars:
  variables: Cepheids -- Stars: oscillations -- Galaxy: bulge --
  Galaxy: disk -- Catalogs}

\vspace*{-5pt}
\Section{Introduction}
The Optical Gravitational Lensing Experiment (OGLE) has discovered the
lar\-gest samples of classical, type~II, and anomalous Cepheids known
to date. The OGLE Collection of Galactic Cepheids (Udalski \etal 2018)
more than doubled the number of known classical Cepheids in the Milky
Way, from about 900 to nearly 2500 objects. This sample was used by
Skowron \etal (2019ab) for a comprehensive examination of the
three-dimensional structure of the Galactic disk -- its warping,
flaring, and the scale height, as well as its radial metallicity
gradient and the recent star formation history. This analysis was
supplemented by Mróz \etal (2019) who used OGLE Cepheids to construct
the most accurate rotation curve of the outer regions of our Galaxy.

The OGLE survey has a long tradition of detecting and studying
variable stars in the Milky Way and nearby galaxies (\eg Udalski \etal
1994, 2015b, 2018, Soszyñski \etal 2008, 2019abc, Pietrukowicz \etal
2013). The OGLE Collection of Variable Stars (OCVS) currently contains
about one million carefully selected and classified variable
stars. These samples are characterized by exceptionally high levels of
completeness and purity and therefore they serve as a framework for
various astrophysical applications. Our collection of Cepheids in the
Galactic bulge and disk (Soszyñski \etal 2017b, Udalski \etal 2018)
was used, among others, to identify the first known double-mode
type~II Cepheids (Smolec \etal 2018), to explore the spatial
distribution and kinematics of the old stellar population in the
central regions of the Milky Way (Braga \etal 2018), to probe the
hydrodynamics of outer envelope and evolutionary status of Cepheids
(Das \etal 2020), as a~training set for machine learning variable star
classification algorithms (D\'ek\'any \etal 2019), and as a
comparative sample for the identification of Cepheids in other regions
of the sky ({\v{Z}}\'ak \etal 2019).

Udalski \etal (2018) presented the first release of the OGLE
Collection of Galactic Cepheids which is a part of the OCVS. This
sample was limited to stars brighter than $I=18$~mag with periods
above 1~d and with the number of photometric data points larger than
100. In the analysis performed by Skowron \etal (2019b), this sample
was extended with 200 classical Cepheids detected in the Galactic
longitude range $20\arcd<l<60\arcd$, however these additional fields
were not searched for type~II and anomalous Cepheids. In this paper,
we supplement the OGLE collection with newly detected classical,
type~II, and anomalous Cepheids in the Milky Way. We also reclassify
several dozen incorrectly categorized variable stars already included
in the OCVS.

\Section{Observations and Data Reduction}
The OCVS was compiled using observations taken by the Warsaw telescope
at Las Campanas Observatory, Chile. The field of view of the OGLE-IV
mosaic CCD camera is 1.4 square degrees, and the pixel scale is
0\zdot\arcs26. In this paper, we use observations collected between
2013 and 2020 during the OGLE Galaxy Variability Survey (GVS) covering
a total of about 3000 square degrees in the Galactic disk and
bulge. Compared to the Udalski's \etal (2018) work, our study covers a
larger area of the sky, in particular, we analyze additional GVS
fields in the outer Galactic bulge and in the northern part of the
disk.

The GVS project has collected {\it I}- and {\it V}-band time series
toward the outer Galactic bulge and disk. The number of available {\it
  I}-band epochs varies across the survey area, with the majority of
the light curves having between 100 and 200 points and a total
baseline of 2--7 yr. Compared to the regular OGLE project, GVS is a
relatively shallow survey, with 25~s and 30~s integration times for
{\it I}-band and {\it V}-band, respectively. The GVS photometry
saturates at $\approx11$~mag, while the faintest stars whose variability
can be effectively studied have $I\approx19.5$~mag. The most recent
observations included in our collection were made on March 13, 2020 --
just before the Las Campanas Observatory suspended operations due to
the COVID-19 pandemic. Details of the instrumentation, image
processing, photometric and astrometric calibrations are discussed by
Udalski \etal (2015a, 2018).

\Section{Variability Search and Classification of Cepheids}
The search for additional Cepheids in the GVS fields was carried out
in the same manner as in the OGLE Collection of Galactic RR~Lyr stars
(Soszyñski \etal 2019c). In brief, we ran the periodicity search
algorithm over all GVS {\it I}-band light curves containing at least
15 data points. We used the {\sc Fnpeaks}
code\footnote{\it http://helas.astro.uni.wroc.pl/deliverables.php?lang=en\&active=fnpeaks}
which provided us the most probable periods with amplitudes and
signal-to-noise ratios. In the second step of the analysis, all
targets with periods shorter than 50~d and the largest amplitudes and
signal-to-noise ratios were fitted with template light curves obtained
from the best-sampled pulsating stars and eclipsing variables observed
by the OGLE survey (typical template light curves of Cepheids and
RR~Lyr stars were shown in the paper by Soszyñski \etal 2019b). In the
final stage of our variable stars' selection procedure, we visually
inspected the light curves pointed by the template fitting algorithm
as the best candidates for pulsating stars and eclipsing binaries. We
isolated candidates for pulsating stars based on the characteristic
attributes of their light curves.

In such a way, we chose a few hundred candidates for Cepheid
variables, not included in the previous edition of the OGLE Collection
of Galactic Cepheids (Udalski \etal 2018). The main diagnostics used
to discriminate between different types of classical pulsators were
their positions in the period -- Fourier parameter diagrams,
especially in the $\log{P}{-}\phi_{21}$ and $\log{P}{-}\phi_{31}$
diagrams, where $\phi_{k1}=\phi_k-k\phi_1$ are phase differences of
the Fourier cosine series fitted to the {\it I}-band light curves
(Simon and Lee 1981).

Since overtone classical Cepheids (including multimode variables) form
a continuity with $\delta$~Sct stars, we had to adopt a boundary
period to separate both types of pulsating variables. Following
previous editions of the OCVS, we adopted the first-overtone pulsation
period of 0.23~d (corresponding to the fundamental-mode period of
about 0.3~d) as a transition between classical Cepheids and
$\delta$~Sct variables. To distinguish between RR~Lyr stars and
(fundamental-mode) type~II Cepheids, we traditionally used the
boundary period of 1~d, although one should be aware that periods of
both classes of pulsating stars partly overlap. Some of the
shortest-period (1.0--1.2~d) BL~Her stars in our collection have light
curves similar to the light curves of the longest-period RRab
variables.

Type II Cepheids have been divided into BL~Her stars (periods in the range
of 1--5~d), W~Vir stars (5--20~d), and RV~Tau stars (periods above
20~d\footnote{For RV~Tau stars and other Cepheids showing period-doubling
  effect, we provide ``single'' periods, \ie intervals between successive
  minima.}). Additionally, we distinguished the subclass of peculiar W~Vir
variables (Soszyñski \etal 2008) and first-overtone BL~Her stars (Soszyñski
\etal 2019a, see Section 5.2).

\hglue-3pt We also updated the OGLE-IV light curves of the previously
published Cepheids in the Milky Way (Soszyñski \etal 2017b, Udalski \etal
2018). The new observations allowed us to correct in some cases our
previous classification. We verified pulsation modes of several Cepheids,
12 other Cepheids were reclassified as a different type of classical
pulsators (for example stars previously classified as classical Cepheids
were recognized as anomalous Cepheids). Finally, 29 objects previously
classified as classical Cepheids and 12 stars classified as type~II
Cepheids were completely removed from the OCVS, because longer light curves
indicated other types of variable stars (usually spotted
variables). Several dozen other Cepheids have been flagged as uncertain in
our collection.

Finally, we matched our collection with the International Variable Star
Index (VSX; Watson \etal 2006), VVV catalog of Cepheids (Braga \etal 2019,
D\'ek\'any \etal 2019), WISE Catalog of Periodic Variable Stars (Chen \etal
2018), Gaia DR2 catalog of Cepheids (Clementini \etal 2019, Ripepi \etal
2019), ASAS-SN catalogs of variable stars (Jurcsik \etal 2018, Jayasinghe
\etal 2020), and ATLAS catalog of pulsating stars (Heinze \etal 2018). We
carefully inspected OGLE light curves of the stars that have not passed our
selection process, but were classified as Cepheids in the above
catalogs. The vast majority of these targets turned out to be non-pulsating
variable stars (eclipsing binaries, spotted variables, irregular variables,
etc.), but we also found several dozen {\it bona fide} Cepheids (mostly
long-period type~II Cepheids and faint classical Cepheids), which were
overlooked during our selection and classification procedure. All these
Cepheids were added to our collection, increasing its completeness.

As a result, we identified 742 Cepheid variables (273 classical
Cepheids, 404 type~II Cepheids, and 65 anomalous Cepheids), not
included in the previous editions of the OCVS. In total, 229 (31\%) of
these objects have been found in at least one of the external catalogs
of variable stars, so the remaining Cepheids are likely new
discoveries. Among others, we found three new triple-mode classical
Cepheids, two unique double-mode 2O/3O Cepheids (Section 5.1),
candidates for type~II Cepheids pulsating solely in the first-overtone
(Section 5.2), and two additional Cepheids in eclipsing binary systems
(Section 5.3).

\Section{The OGLE Collection of Galactic Cepheids}
The newly identified Cepheids have been added to the OGLE Collection of
Galactic Cepheids. The collection contains now 1974 classical Cepheids,
1625 type~II Cepheids, and 119 anomalous Cepheids in the Galactic bulge and
disk. The exact numbers of Cepheids belonging to different subtypes are
provided in Table~1. The entire collection is available {\it via} the WWW
interface or anonymous FTP sites:
\begin{center}
{\it http://ogle.astrouw.edu.pl}\\
{\it ftp://ftp.astrouw.edu.pl/ogle/ogle4/OCVS/gd/cep/}\\
{\it ftp://ftp.astrouw.edu.pl/ogle/ogle4/OCVS/blg/cep/}\\
{\it ftp://ftp.astrouw.edu.pl/ogle/ogle4/OCVS/gd/t2cep/}\\
{\it ftp://ftp.astrouw.edu.pl/ogle/ogle4/OCVS/blg/t2cep/}\\
{\it ftp://ftp.astrouw.edu.pl/ogle/ogle4/OCVS/gal/acep/}\\
\end{center}

\MakeTableee{l@{\hspace{10pt}}
l@{\hspace{10pt}}
r@{\hspace{10pt}}
r@{\hspace{10pt}}
r@{\hspace{5pt}}}{12.5cm}{Number of Galactic Cepheids in the OCVS}
{\hline
\noalign{\vskip3pt}
\multicolumn{1}{c}{Type}
& Subtype
&\multicolumn{1}{c}{Bulge}
&\multicolumn{1}{c}{Disk/Halo}
&\multicolumn{1}{r}{Total}\\
\noalign{\vskip3pt}
\hline
\noalign{\vskip3pt}
Classical Cepheids & all          &  189 & 1784 & 1973 \\
                   & F            &   98 & 1103 & 1201 \\
                   & 1O           &   56 &  495 &  551 \\
                   & F/1O         &   12 &   45 &   57 \\
                   & 1O/2O        &   19 &  133 &  152 \\
                   & 2O/3O        &    1 &    1 &    2 \\
                   & F/1O/2O      &    0 &    2 &    2 \\
                   & 1O/2O/3O     &    3 &    5 &    8 \\
\noalign{\vskip3pt}
\hline
\noalign{\vskip3pt}
Type II Cepheids   & all          & 1320 &  306 & 1626 \\
                   & BL~Her (F)   &  541 &  120 &  661 \\
                   & BL~Her (1O)  &    1 &    2 &    3 \\
                   & BL~Her (F/1O)&    3 &    2 &    5 \\
                   & W~Vir        &  494 &   98 &  592 \\
                   & pec. W~Vir   &   45 &   20 &   65 \\
                   & RV~Tau       &  236 &   64 &  300 \\
\noalign{\vskip3pt}
\hline
\noalign{\vskip3pt}
Anomalous Cepheids & all          &   65 &   54 &  119 \\
                   & F            &   46 &   35 &   81 \\
                   & 1O           &   18 &   19 &   37 \\
                   & F/1O/2O      &    1 &    0 &    1 \\
\noalign{\vskip3pt}
\hline}

For each star, we provide its equatorial coordinates (J2000.0), pulsation
periods, intensity-averaged mean magnitudes in the {\it I}- and {\it
  V}-bands, epochs of the maximum light, peak-to-peak {\it I}-band
amplitudes, and Fourier parameters derived from the {\it I}-band light
curves. The photometric time series in the {\it I}- and {\it V}-bands (if
available) and finding charts are also provided for each Cepheid.
\begin{landscape}
\begin{figure}[p]
\centerline{\includegraphics[bb = 80 20 495 800, clip, width=11cm, angle=-90]{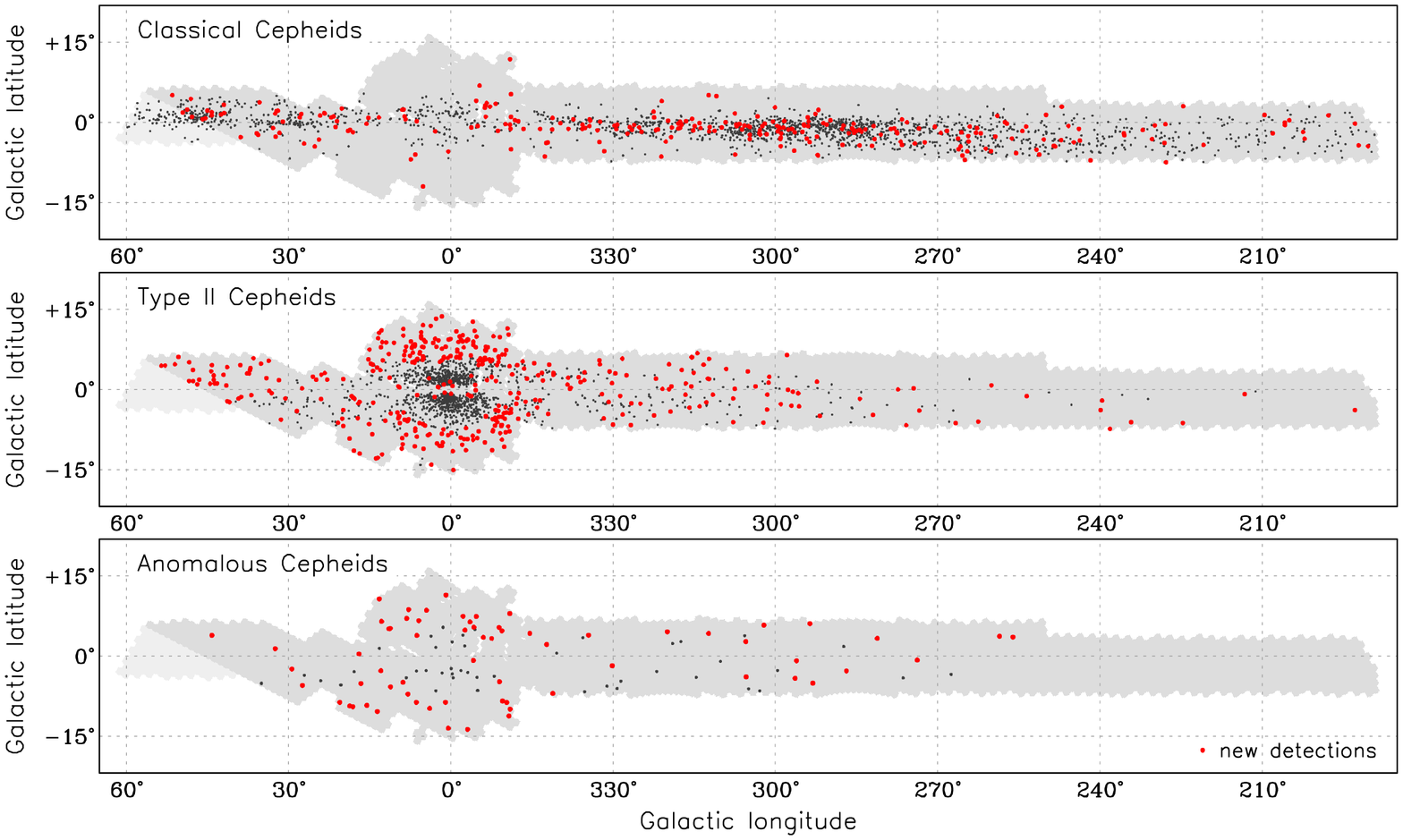}}
\vskip9pt
\FigCap{On-sky distribution of classical Cepheids ({\it upper panel}),
  type~II Cepheids ({\it middle panel}), and anomalous Cepheids ({\it
    lower panel}) in Galactic coordinates. Dark gray dots mark objects
  included in the previous edition of the OCVS, while red dots indicate the
  newly detected and reclassified Cepheids. The gray area shows the OGLE
  footprint in the Galactic bulge and disk (the lighter shade indicates an
  area not included in this investigation because of an insufficient number
  of observations).}
\end{figure}
\end{landscape}

The light curves of previously published Cepheids have been supplemented
with the newest OGLE-IV observations. These extended photometric data were
used to update the pulsation periods and other observational
quantities. Note that a fraction of type~II Cepheids, in particular W~Vir
stars, exhibits variable periods which makes it impossible to phase their
OGLE-IV light curves using constant periods. To recompute the periods we
used the {\sc Tatry} code, which implements the multiharmonic analysis of
variance algorithm (Schwarzenberg-Czerny 1996).

The updated list of all known Galactic classical Cepheids is available from
the OGLE Internet archive:
\begin{center}
{\it ftp://ogle.astrouw.edu.pl/ogle/ogle4/OCVS/allGalCep.listID}\\
\end{center}

The celestial distributions of classical, type~II and anomalous Cepheids
are shown in Fig.~1. The distribution of classical Cepheids (upper panel of
Fig.~1) clearly reveals the warping of the Galactic disk (Skowron \etal
2019ab): most of the Cepheids are located below the Galactic plane at the
Galactic longitudes $l<330\arcd$ and above the Galactic plane at
$l>30\arcd$. The width of the distribution grows with increasing distance
from the center of the Galaxy, which provides evidence for the disk
flaring.

The middle panel of Fig.~1 shows that type~II Cepheids are strongly
concentrated toward the Galactic center, which resembles the spatial
distribution of RR~Lyr stars (Soszyñski \etal 2019c). We identified more
than 30 type~II Cepheids per square degree in the most populated fields in
the central Galactic bulge, but their surface density drops to below 0.5
objects per square degree in the outer bulge, while close to the Galactic
anticenter there are only a few type~II Cepheids in the area of several
hundred square degrees. Anomalous Cepheids (lower panel of Fig.~1) also
exhibit a Galactocentric distribution, but the gradient of their surface
density is much less steep than that for type~II Cepheids. It is worth
noting that anomalous Cepheids in the Large (LMC) and Small Magellanic
Cloud (SMC) show similar smooth distributions around the centers of these
galaxies (Soszyñski \etal 2017a).

\Section{Discussion}
\Subsection{Multi-Mode Cepheids}
\vskip3pt
{\it Classical Cepheids}
\vskip5pt

\begin{figure}[p]
\includegraphics[bb = 30 50 550 740, clip, width=13.0cm]{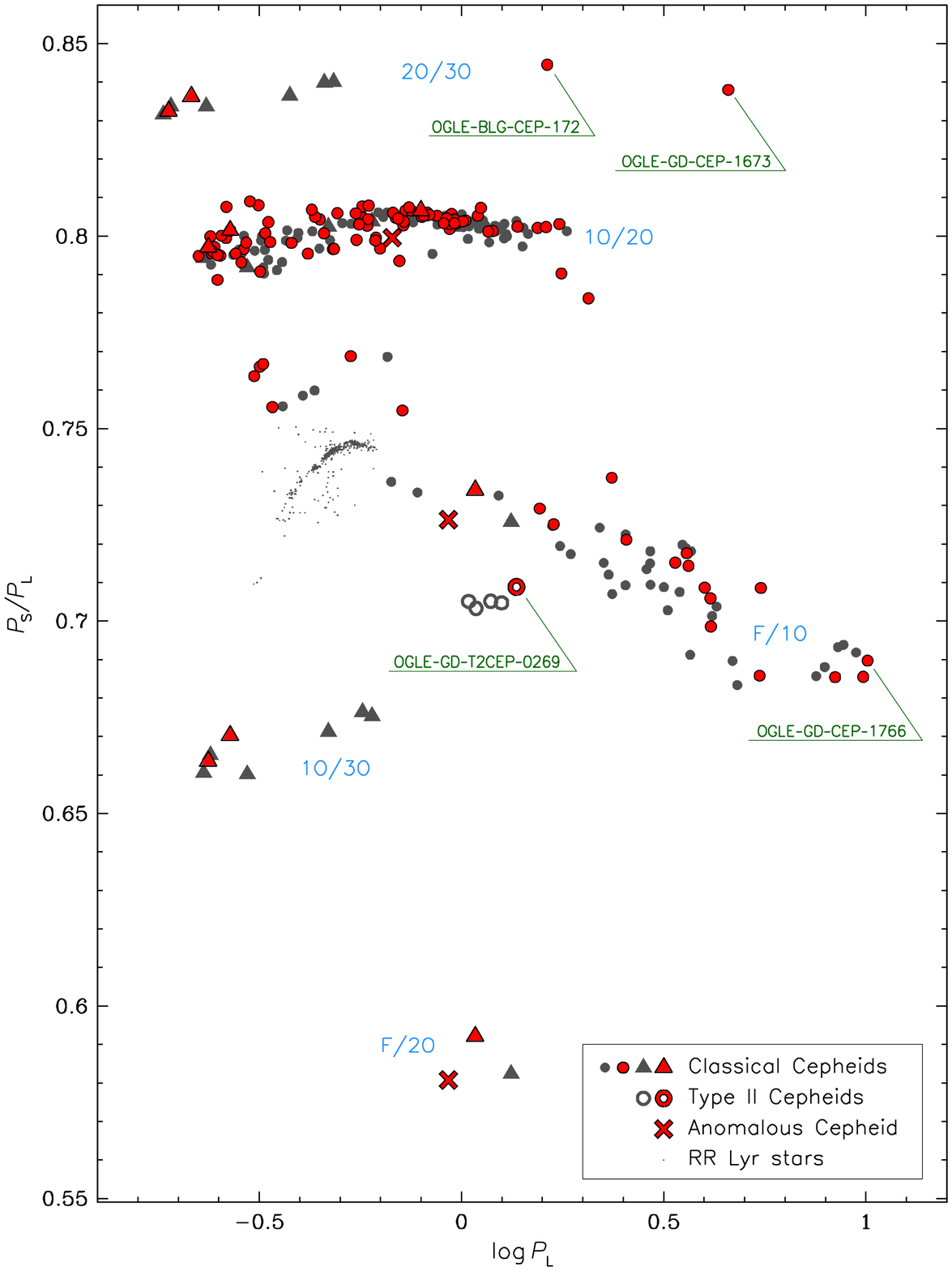}
\FigCap{Petersen diagram for multi-mode Cepheids in the Milky Way. Filled
  circles mark double-mode classical Cepheids, triangles -- triple-mode
  classical Cepheids (three points per star), empty circles -- double-mode
  type~II Cepheids, crosses -- a triple-mode anomalous Cepheid
  OGLE-GAL-ACEP-091. Gray and red symbols present previously known Cepheids
  (Soszyñski \etal 2017b, Udalski \etal 2018) and new detections,
  respectively.}
\end{figure}

Stars revealing the simultaneous presence of two or three radial modes
constitute a significant fraction of the total sample of classical
Cepheids. In the virtually complete OGLE collection of classical Cepheids
in the Magellanic Clouds (Soszyñski \etal 2019b), multimode oscillations
occur in about 6\% variables in the SMC and about 9\% Cepheids in the
LMC. In our sample of Galactic classical Cepheids, as many as 11\% objects
are double- or triple-mode pulsators, so the incidence rate of the beat
Cepheids seems to be higher in the environments of higher
metallicity. Interestingly, the opposite relationship was found for RR~Lyr
stars: the incidence rates of double-mode RR~Lyr stars are equal to 10\%,
5\%, and 0.5\% for the SMC, LMC, and Galactic bulge samples, respectively
(Soszyñski \etal 2019c).

Fig.~2 presents the Petersen diagram (shorter to longer period ratios
plotted against logarithmic longer periods) for multimode classical
pulsators in the Milky Way. Our new detections are marked with red circles
and triangles for double- and triple-mode variables, respectively.

\begin{figure}[t]
\vglue-3mm
\includegraphics[bb = 35 300 570 760, clip, width=12.8cm]{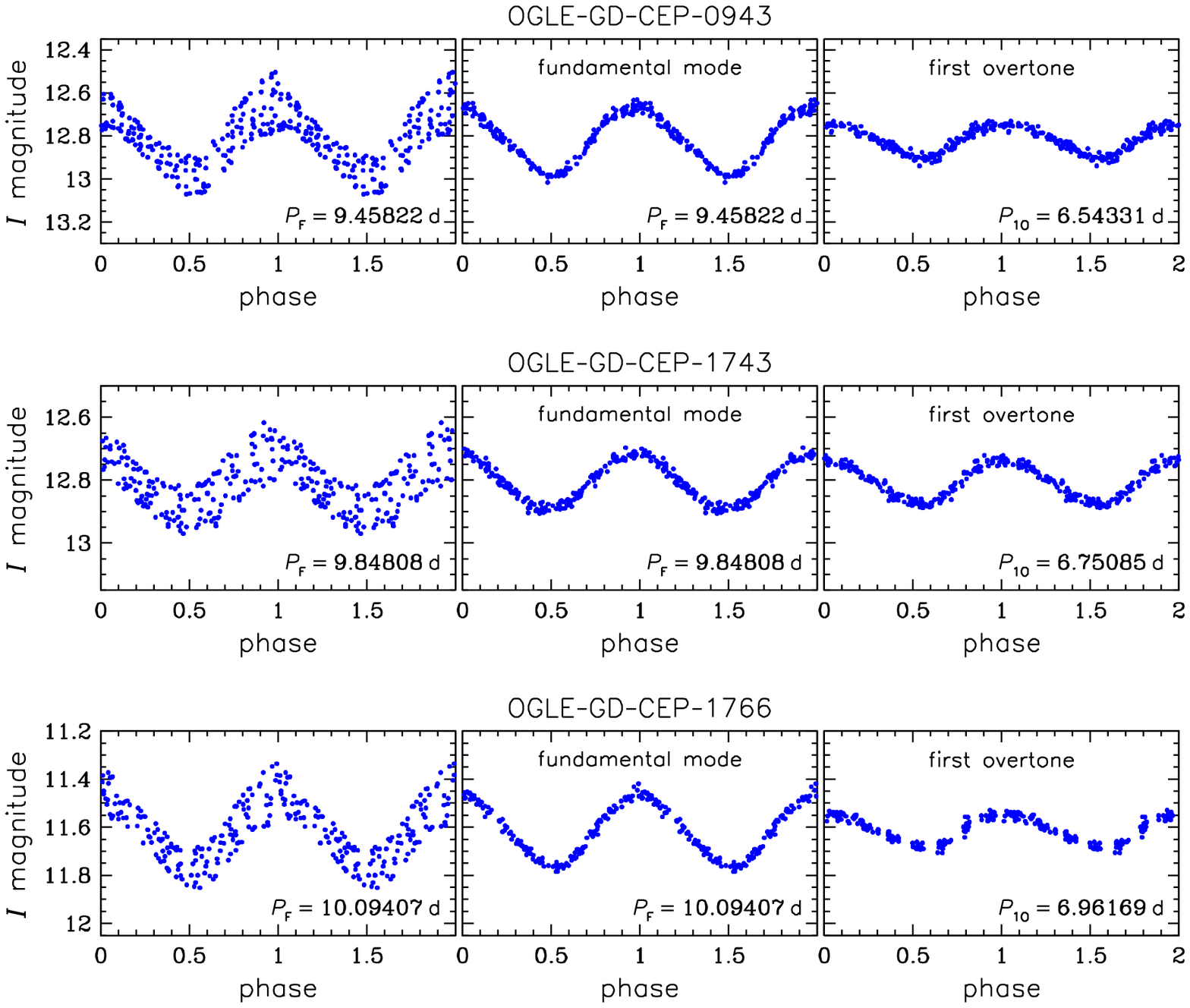}
\vskip-3pt
\FigCap{Light curves of the longest-period double-mode F/1O
classical Cepheids in the Milky Way. {\it Left panels} show original
light curves folded with the fundamental-mode periods. {\it Middle}
and {\it right panels} present light curves of the fundamental and
first-overtone modes, respectively, prewhitened for the other radial
modes and for the linear combination terms.}
\end{figure}

First of all, we draw the reader's attention to the isolated sub-group of
eight long-period double-mode F/1O classical Cepheids with the
fundamental-mode periods in the range of 7.5--10~d and period ratios of
about 0.69. The light curves of three the longest-period stars from this
group are shown in Fig.~3. From this group, OGLE-GD-CEP-1766, with $P_{\rm
  F}=10.0940$~d, is probably the longest-period beat Cepheid known in the
Universe\footnote{Smolec (2017) found in the OCVS an even longer-period
  candidate for double-mode Cepheid -- OGLE-SMC-CEP-0387 with $P_{\rm
    F}\approx13.1$~d. However, the identification of the first-overtone
  mode in this star is very uncertain.}. Two similar long-period
double-mode F/1O Cepheids were discovered by Poleski (2013) in M31. The
existence of such variables was theoretically predicted by Dziembowski and
Kov\'acs (1984) and Buchler (2009), who claimed that the F/1O bimodality in
the vicinity of the $P_{\rm F}=10$~d period is induced by the $P_{\rm
  F}=2P_{\rm 2O}$ resonance. Indeed, at least some of our targets show
strong $2f_{\rm F}+f_{\rm 1O}$ combination frequencies, which can be
interpreted as $f_{\rm 2O}+f_{\rm 1O}$ frequencies.

\begin{figure}[t]
\includegraphics[bb = 35 460 570 760, clip, width=12.7cm]{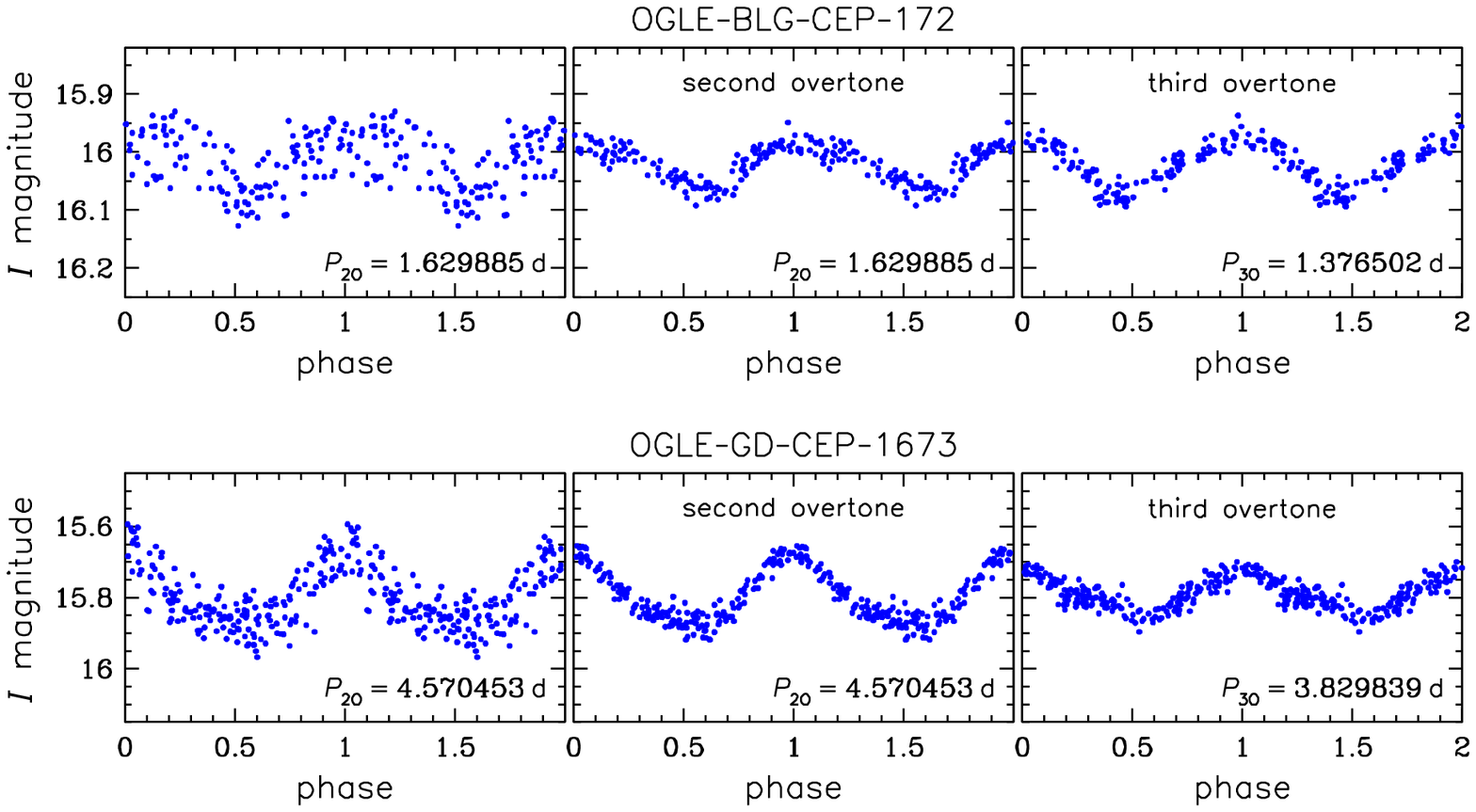}
\FigCap{Light curves of double-mode 2O/3O classical Cepheids in the
Milky Way. {\it Left panels} show original light curves folded with
the second-overtone periods. {\it Middle} and {\it right panels}
present disentangled light curves of the second- and third-overtone
modes, respectively.}
\end{figure}

\MakeTableee{l@{\hspace{8pt}}
c@{\hspace{6pt}}
c@{\hspace{8pt}}
c@{\hspace{8pt}}   
c@{\hspace{8pt}}
c@{\hspace{8pt}}
c@{\hspace{3pt}}}
{12.5cm}{Double-mode 2O/3O classical Cepheids in the Milky Way}
{\hline
\noalign{\vskip3pt}
\multicolumn{1}{c}{Identifier}
& R.A.
& Dec.
& $P_{\rm 2O}$
& $P_{\rm 3O}$
& $\langle{I}\rangle$
& $\langle{V}\rangle$ \\
& [J2000.0]
& [J2000.0]
& [d]
& [d]
& [mag]
& [mag] \\
\noalign{\vskip3pt}
\hline
\noalign{\vskip3pt}  
OGLE-GD-CEP-1673 & 10\uph43\upm58\zdot\ups84 & $-59\arcd34\arcm28\zdot\arcs0$ & 4.570453 & 3.829839 & 15.791 & 17.929 \\
OGLE-BLG-CEP-172 & 17\uph10\upm20\zdot\ups43 & $-33\arcd52\arcm55\zdot\arcs9$ & 1.629885 & 1.376502 & 16.008 &   --   \\
\noalign{\vskip3pt}
\hline}

Two other unique objects in our collection -- OGLE-GD-CEP-1673 and
OGLE-BLG-CEP-172 -- are double-mode Cepheids pulsating in the second and
third overtone modes. Table~2 lists their basic parameters, while their
disentangled light curves are shown in Fig.~4. The period ratios (around
0.84) of our 2O/3O beat Cepheids place them at the top of the Petersen
diagram (Fig.~2). So far, only one double-mode 2O/3O classical Cepheid was
recognized: an LMC variable OGLE-LMC-CEP-3987 (Soszyñski \etal 2015),
however this object has much shorter periods ($P_{\rm 2O}\approx 0.5342$~d,
$P_{\rm 3O}\approx 0.4483$~d) than our Galactic Cepheids. We believe that
OGLE-GD-CEP-1673 and OGLE-BLG-CEP-172 will be important targets of
asteroseismological research.

\vspace*{5mm}
{\it Type II and Anomalous Cepheids}
\vspace*{2mm}

In contrast to classical Cepheids, double-mode type~II Cepheids are
extremely rare objects. Smolec \etal (2018) identified the first two BL~Her
variables with the fundamental and first-overtone modes simultaneously
excited. Shortly after, Udalski \etal (2018) reported the discovery of two
other stars belonging to the same class. The four known beat type~II
Cepheids constitute a very homogeneous group: their fundamental-mode
periods $P_{\rm F}$ range from 1.04 to 1.26~d, their period ratios $P_{\rm
1O}/P_{\rm F}$ are in the range of 0.7033--0.7051, the peak-to-peak {\it
I}-band amplitudes of the fundamental-mode light curves are equal to
about 0.4 mag and are several times larger than the amplitudes of the
first-overtone mode. Finally, the light curve shape is also remarkably
similar in these four stars.

\begin{figure}[h]
\includegraphics[bb = 35 610 570 760, clip, width=12.8cm]{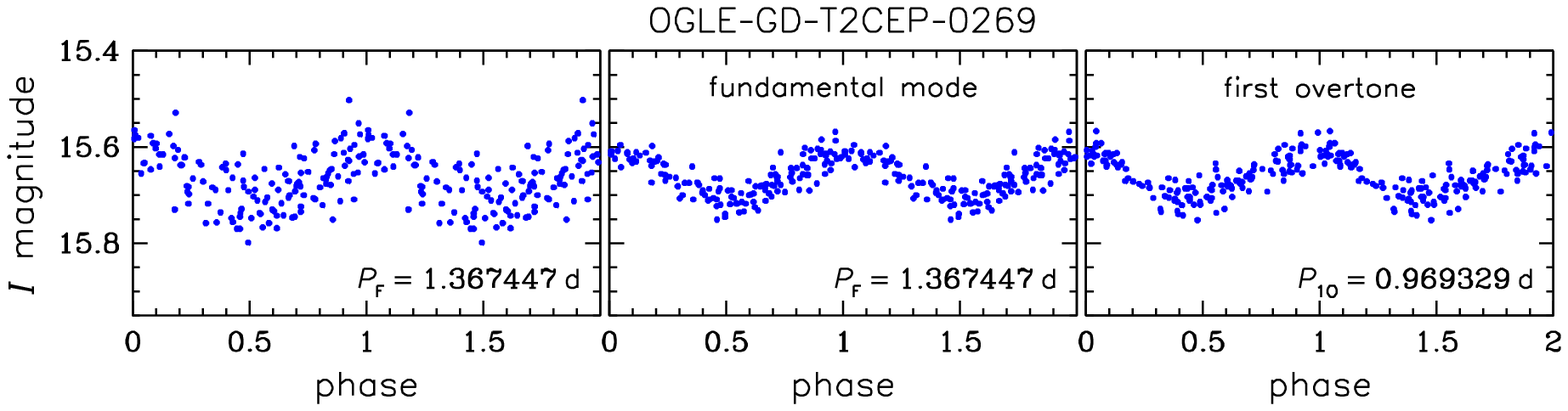}
\FigCap{Light curve of a newly detected double-mode F/1O type~II Cepheid
OGLE-GD-T2CEP-0269. {\it Left panel} shows original light curves folded with
the fundamental-mode period. {\it Middle} and {\it right panels}
present disentangled light curves of the fundamental and first-overtone 
modes, respectively.}
\end{figure}

\MakeTableee{l@{\hspace{8pt}}
c@{\hspace{6pt}}
c@{\hspace{8pt}}
c@{\hspace{8pt}}   
c@{\hspace{8pt}}
c@{\hspace{8pt}}
c@{\hspace{3pt}}}
{12.5cm}{Double-mode type~II Cepheids (BL~Her stars) in the Milky Way}
{\hline
\noalign{\vskip3pt}
\multicolumn{1}{c}{Identifier}
& R.A.
& Dec.
& $P_{\rm F}$
& $P_{\rm 1O}$
& $\langle{I}\rangle$
& $\langle{V}\rangle$ \\
& [J2000.0]
& [J2000.0]
& [d]
& [d]
& [mag]
& [mag] \\
\noalign{\vskip3pt}
\hline
\noalign{\vskip3pt}  
OGLE-GD-T2CEP-0045  & 13\uph27\upm58\zdot\ups53 & $-67\arcd25\arcm01\zdot\arcs5$ & 1.086491 & 0.764100 & 13.899 & 14.857 \\
OGLE-GD-T2CEP-0269  & 16\uph50\upm18\zdot\ups84 & $-41\arcd51\arcm12\zdot\arcs8$ & 1.367447 & 0.969329 & 15.666 & 17.624 \\
OGLE-BLG-T2CEP-0209 & 17\uph57\upm11\zdot\ups87 & $-28\arcd51\arcm49\zdot\arcs5$ & 1.181284 & 0.832978 & 15.445 & 17.279 \\
OGLE-BLG-T2CEP-0749 & 17\uph50\upm05\zdot\ups11 & $-32\arcd16\arcm17\zdot\arcs2$ & 1.041577 & 0.734419 & 17.529 & 20.852 \\
OGLE-BLG-T2CEP-1041 & 18\uph12\upm35\zdot\ups05 & $-15\arcd31\arcm16\zdot\arcs0$ & 1.256607 & 0.885549 & 15.718 &   --   \\
\noalign{\vskip3pt}
\hline}

In this work, we supplement the list of double-mode BL~Her stars with one
more object: OGLE-GD-T2CEP-0269. The pulsation periods of this star
($P_{\rm F}=1.36745$~d, $P_{\rm 1O}/P_{\rm F}=0.7089$) are slightly longer
than the periods of other beat type~II Cepheids, but the position of
OGLE-GD-T2CEP-0269 in the Petersen diagram (Fig.~2) indicates that it is a
plausible member of the same class of double-mode pulsators. Fig.~5 shows
disentangled light curves of OGLE-GD-T2CEP-0269. The amplitudes of the two
modes are nearly equal to each other, which distinguishes our new detection
from the other double-mode BL~Her stars. Table~3 summarizes all known
double-mode type~II Cepheids in the Milky Way.

Finally, we also identified the first known triple-mode anomalous Cepheid
-- OGLE-GAL-ACEP-091. Its thorough analysis will be presented in a
forthcoming paper.

\Subsection{Single-Mode First-Overtone Type II Cepheids}
Until recently, all known type~II Cepheids were pure fundamental-mode
pulsators. The two double-mode BL~Her stars discovered by Smolec \etal
(2018) were the first known type~II Cepheids with the first-overtone mode
excited. Soon after, Soszyñski \etal (2019a) reported the discovery of two
stars in the LMC that are strong candidates to be single-mode
first-overtone type~II Cepheids. Soszyñski's \etal (2019a) classification
is based mostly on the position of these stars in the period--luminosity
diagram and their characteristic light curve morphology. Unfortunately, the
former criterion cannot be easily applied to Cepheids in the Milky Way,
because distances to individual targets are not {\it a priori} known. Thus,
it is impossible to place these stars in the period--absolute luminosity
diagram. Taking this limitation into consideration, we performed a search
for Galactic overtone type~II Cepheids based on their light curve
characteristics.

The two candidates for overtone BL~Her stars in the LMC --
OGLE-LMC-T2CEP-290 and OGLE-LMC-T2CEP-291 -- pulsate with periods of about
0.93~d and 0.82~d, respectively. Their light curves (see Fig.~2 in Soszyñski
\etal 2019a) are characterized by round minima (the feature typical for
overtone pulsators) and sharp maxima. Although overtone classical,
anomalous and type~II Cepheids have relatively similar light curves, a
careful analysis of the Fourier parameters $\phi_{21}$, $\phi_{31}$,
$R_{21}$, $R_{31}$ can make it possible to distinguish these three types of
classical pulsators.
 
\begin{figure}[h]
\includegraphics[bb = 30 375 570 760, clip, width=11.6cm]{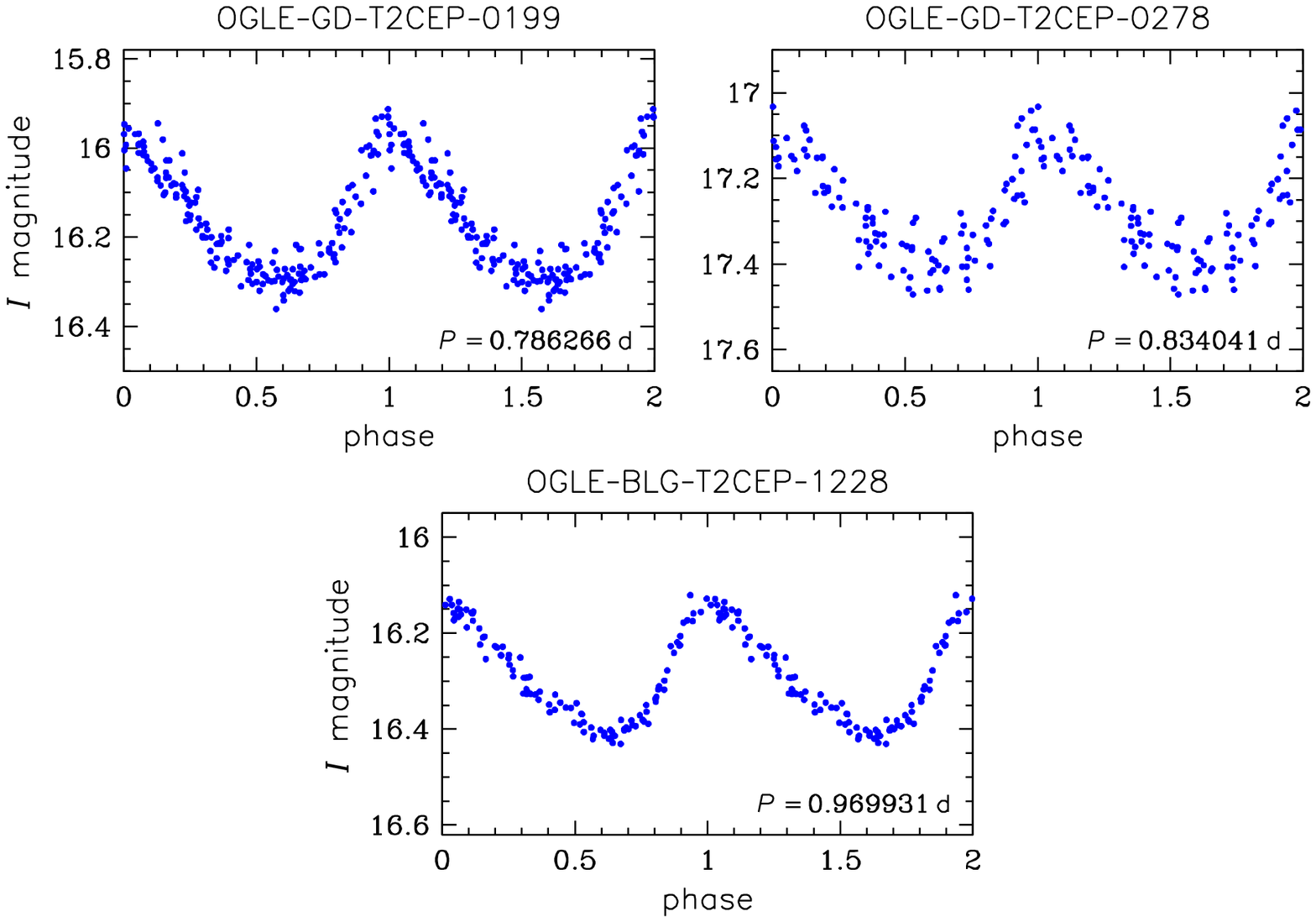}
\FigCap{Light curves of candidates for single-mode first-overtone type~II
Cepheids (BL~Her stars) in the Milky Way.}
\end{figure}

We restricted our collection of Cepheids in the Milky Way to single-mode
pulsators with periods in the range 0.7--1~d, and we conducted a search for
light curves with round minima and sharp maxima -- resembling those of
OGLE-LMC-T2CEP-290 and OGLE-LMC-T2CEP-291. The result of our search is
displayed in Fig.~6, showing the light curves of three convincing
candidates for the Galactic first-overtone BL~Her stars. Their basic
parameters (identifiers, J2000.0 equatorial coordinates, periods, mean
magnitudes in the {\it I} and {\it V} bands, and {\it I}-band amplitudes)
are listed in Table~4. We emphasize that our candidates for first-overtone
type~II Cepheids in the Milky Way are not as certain as the two candidates
found by Soszyñski \etal (2019a) in the LMC, because we currently cannot
verify their positions on the period--luminosity diagram.

\MakeTable{l@{\hspace{8pt}}
c@{\hspace{6pt}}
c@{\hspace{8pt}}
c@{\hspace{8pt}}   
c@{\hspace{8pt}}
c@{\hspace{8pt}}
c@{\hspace{3pt}}}
{12.5cm}{Candidates for single-mode first-overtone type~II Cepheids (BL~Her stars).}
{\hline
\noalign{\vskip3pt}
\multicolumn{1}{c}{Identifier}
& R.A.
& Dec.
& $P_{\rm 1O}$
& $\langle{I}\rangle$
& $\langle{V}\rangle$
& $A(I)$ \\
& [J2000.0]
& [J2000.0]
& [d]
& [mag]
& [mag]
& [mag] \\
\noalign{\vskip3pt}
\hline
\noalign{\vskip3pt}  
OGLE-GD-T2CEP-0199  & 11\uph54\upm27\zdot\ups70 & $-65\arcd13\arcm49\zdot\arcs5$ & 0.786266 & 16.154 & 18.433 & 0.332 \\
OGLE-GD-T2CEP-0278  & 18\uph29\upm08\zdot\ups06 & $-05\arcd28\arcm27\zdot\arcs9$ & 0.834041 & 17.264 &   --   & 0.311 \\
OGLE-BLG-T2CEP-1228 & 17\uph42\upm20\zdot\ups75 & $-20\arcd36\arcm34\zdot\arcs4$ & 0.969931 & 16.285 &   --   & 0.274 \\
\noalign{\vskip3pt}
\hline}

\begin{figure}[p]
\includegraphics[bb = 35 50 570 740, clip, width=13.0cm]{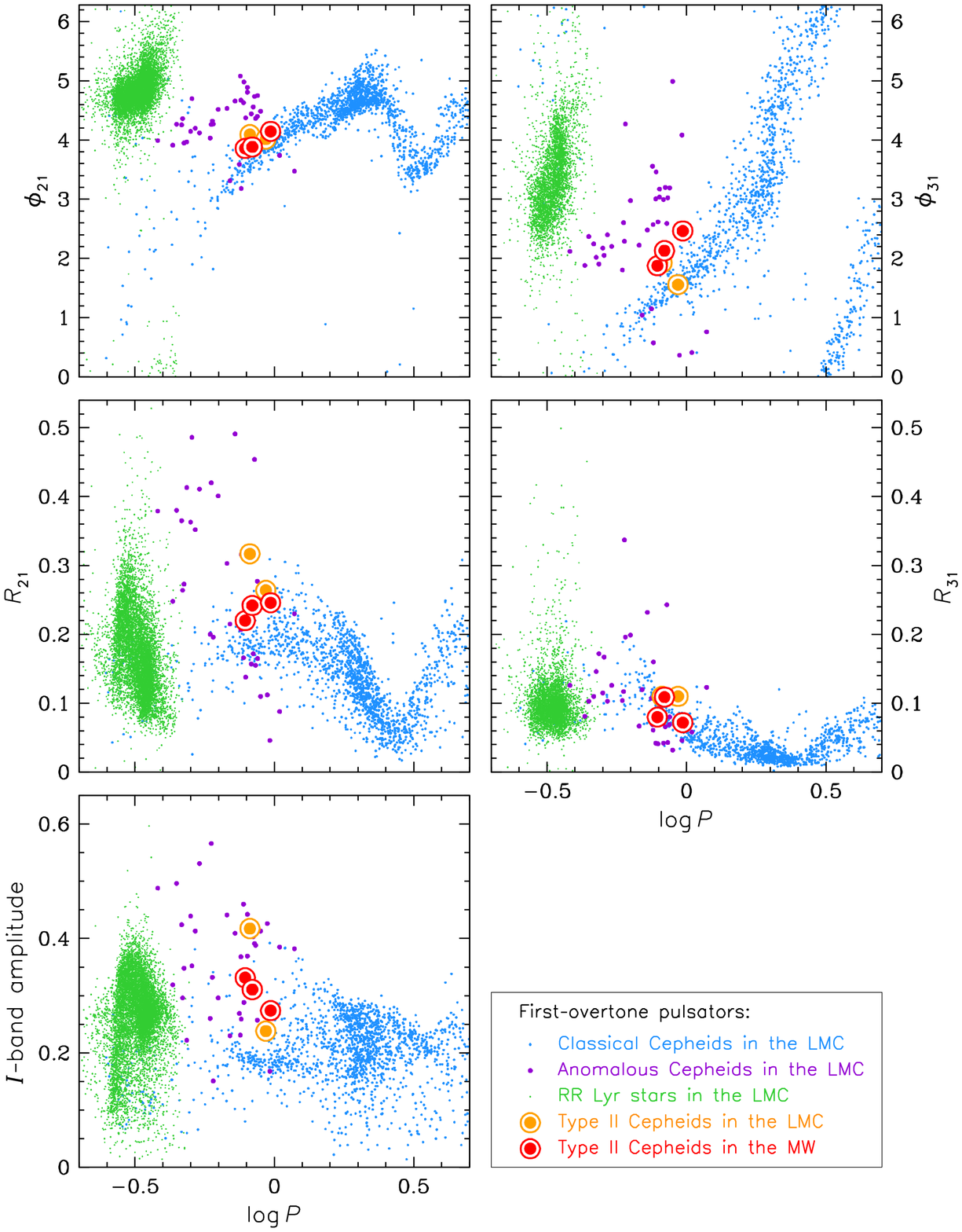}
\FigCap{Fourier coefficients $\phi_{21}$, $\phi_{31}$, $R_{21}$,
$R_{31}$, and {\it I}-band peak-to-peak amplitudes as a function of
periods for Cepheids and RR~Lyr stars pulsating in the first-overtone
mode. Blue, violet, and green points indicate classical Cepheids,
anomalous Cepheids, and RRc stars in the LMC, respectively. Orange
circles represent two first-overtone type~II Cepheids identified by
Soszyñski \etal (2019a) in the LMC. Red circles mark three candidates
for first-overtone type~II Cepheids in the Milky Way.}
\end{figure}

\newpage 
Our candidates for overtone type~II Cepheids are relatively faint in the
shallow GVS, which makes their time series quite noisy
(Fig.~6). Nevertheless, we have determined the Fourier coefficients and
peak-to-peak amplitudes of their light curves and plotted these parameters
against the pulsation periods in Fig.~7. For comparison, the same
quantities derived for the first-overtone classical pulsators (RRc stars,
classical Cepheids, anomalous Cepheids, and type~II Cepheids) in the LMC
are shown in these diagrams. Note that our three candidates for Galactic
overtone type~II Cepheids have Fourier parameters very similar to the
parameters of their siblings in the LMC. At the same time, their Fourier
parameters differ from the parameters of overtone anomalous Cepheids with
similar periods and have larger amplitudes than the bulk of overtone
classical Cepheids with similar periods. Further studies, including a
dedicated spectroscopic observations, should shed more light on the nature
of our candidates for overtone BL~Her stars.

\Subsection{Cepheids in Eclipsing Binary Systems}
Eclipsing binary systems with pulsating components are very rare, but
important astrophysical targets. Such objects give us an opportunity to
directly and very precisely measure stellar parameters like masses,
absolute luminosities, and radii. Several classical Cepheids in eclipsing
binary systems have been discovered in the Magellanic Clouds (\eg Udalski
\etal 2015b). These objects have been the subject of extensive studies by
the Araucaria project (\eg Pietrzyñski \etal 2010). Until recently, no
candidates for the eclipsing classical Cepheid were known in our Galaxy.

Udalski \etal (2018) reported the discovery of two Galactic classical
Cepheids with additional eclipsing modulations. However, in this paper we
reclassify one of these stars (OGLE-GD-CEP-0069) as a peculiar W~Vir star
and we move this object to the collection of type~II Cepheids (its new
designation is OGLE-GD-T2CEP-0178). It is known that a large fraction of
peculiar W~Vir stars (Soszyñski \etal 2008) are members of binary systems
-- it can be even supposed that the binarity is necessary for the
occurrence of pulsations in these variables. The morphology of both,
pulsation and eclipsing, light curve components of OGLE-GD-T2CEP-0178 are
more compatible with the variability of a peculiar W~Vir star in an
eclipsing binary system (\cf Fig. 5 in Soszyñski \etal 2018) than with a
classical Cepheid.

In turn, OGLE-GD-CEP-0465 -- the other pulsator revealing simultaneous
eclip\-sing variability presented by Udalski \etal (2018) -- is a certain
fundamental-mode classical Cepheid. This is currently the only known
Galactic classical Cepheid which can be a member of an eclipsing
system. Spectroscopic observations of this object should answer the
question whether this is a binary system hosting $\delta$~Cep star or
rather the Cepheid and eclipsing variable are physically unrelated blend
observed by chance along the same line-of-sight.

\begin{figure}[h]
\includegraphics[bb = 35 460 570 760, clip, width=12.9cm]{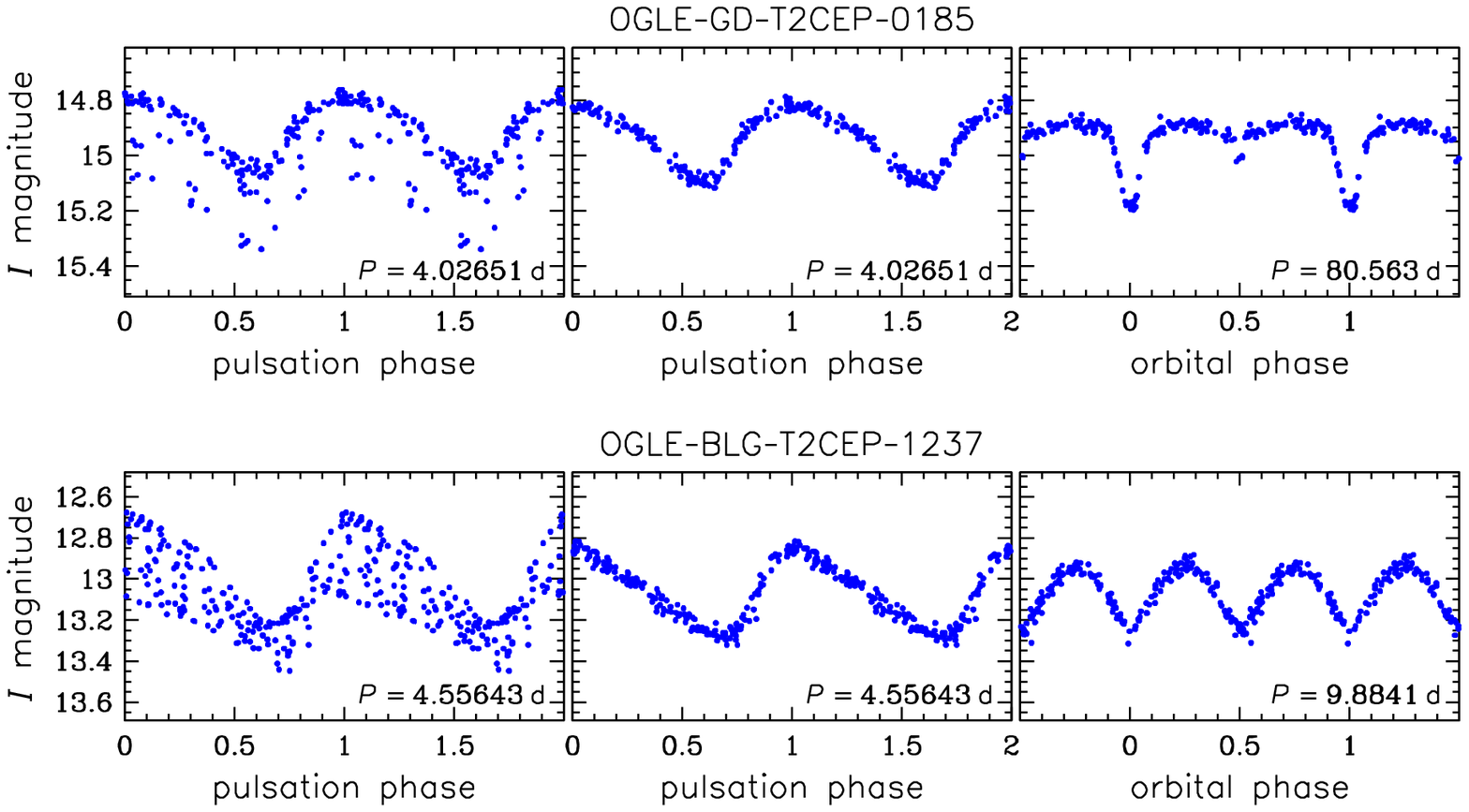}
\FigCap{Light curves of newly detected type~II Cepheids with
additional eclipsing variability. {\it Left panels} show original
light curves folded with the pulsation periods. {\it Middle} and {\it
right panels} present pulsation and eclipsing light curves,
respectively, after subtracting the other component.}
\end{figure}

In this paper, we supplement the short list of Galactic Cepheids in
eclipsing binary systems with two new objects of that type:
OGLE-GD-T2CEP-0185 and OGLE-BLG-T2CEP-1237. Both stars are classified as
peculiar W~Vir stars, although their pulsation periods are shorter than 5~d
(which we adopted as a transition between BL~Her and W~Vir stars). The
disentangled light curves of both objects are presented in Fig.~8. In the
case of OGLE-GD-T2CEP-0185, different widths of the primary and secondary
eclipses indicate the presence of a disk in the system. The same effect is
visible, among others, in OGLE-GD-T2CEP-0178 and OGLE-LMC-T2CEP-211, the
latter examined in detail by Pilecki \etal (2018). The shape of the
eclipsing component of OGLE-BLG-T2CEP-1237 indicates that it is a close
binary system. It is worth noting that additional frequencies that are
linear combinations of the pulsation and orbital periods are visible in the
pre-whitened light curve of this object. This feature proves that the
pulsating star is a real member of the system. Such combination of
frequencies reflect complex oscillations of stars which are distorted by
tidal interactions from their companions.

Table~5 lists all known Galactic Cepheids revealing eclipsing modulation. 

\MakeTable{l@{\hspace{8pt}}
c@{\hspace{6pt}}
c@{\hspace{8pt}}
c@{\hspace{8pt}}
c@{\hspace{8pt}}
c@{\hspace{8pt}}
c@{\hspace{3pt}}}
{12.5cm}{Milky Way Cepheids showing additional eclipsing variability}
{\hline
\noalign{\vskip3pt}
\multicolumn{1}{c}{Identifier}
& R.A.
& Dec.
& $P_{\rm puls}$
& $P_{\rm ecl}$
& $\langle{I}\rangle$
& $\langle{V}\rangle$ \\
& [J2000.0]
& [J2000.0]
& [d]
& [d]
& [mag]
& [mag] \\
\noalign{\vskip3pt}
\hline
\noalign{\vskip3pt}
\multicolumn{7}{c}{Classical Cepheid}\\
\noalign{\vskip3pt}
\hline
\noalign{\vskip3pt}
OGLE-GD-CEP-0465   & 10\uph14\upm30\zdot\ups19 & $-59\arcd13\arcm36\zdot\arcs3$ & 6.605641 & 193.83 & 14.411 & 17.127 \\
\noalign{\vskip3pt}
\hline
\noalign{\vskip3pt}
\multicolumn{7}{c}{Type II Cepheids}\\
\noalign{\vskip3pt}
\hline
\noalign{\vskip3pt}
OGLE-GD-T2CEP-0178  & 06\uph48\upm58\zdot\ups05 & $-00\arcd37\arcm30\zdot\arcs8$ & 3.833057 & 81.640 & 10.855 & 11.832 \\
OGLE-GD-T2CEP-0185  & 08\uph16\upm54\zdot\ups77 & $-46\arcd09\arcm44\zdot\arcs6$ & 4.026511 & 80.563 & 14.931 & 16.229 \\
OGLE-BLG-T2CEP-0674 & 17\uph45\upm00\zdot\ups28 & $-27\arcd15\arcm08\zdot\arcs3$ & 6.985900 & 714    & 16.739 & 21.395 \\
OGLE-BLG-T2CEP-1237 & 17\uph47\upm01\zdot\ups12 & $-24\arcd23\arcm31\zdot\arcs1$ & 4.556427 & 9.8841 & 13.058 & 15.249 \\
\noalign{\vskip3pt}
\hline}

\Subsection{Cepheids in Globular Clusters}
We matched our collection of Galactic Cepheids with the catalog of globular
clusters in the Milky Way compiled by Harris (2010). We searched for stars
located in the sky not farther than one tidal radius from the center of a
cluster. Our collection contains exactly 60 Galactic type~II Cepheids which
could potentially be members of 25 globular clusters. The list of these
stars is provided in the file {\sf gc.dat} in the FTP site of the
collection. Of course, some of these stars are probably not physically
associated with the clusters -- their positional coincidence are
accidental. However, the discrimination between cluster members and field
variables is beyond the scope of this paper. The most Cepheid-rich globular
clusters in our fields are Terzan~1, NGC~6273 (M19), and NGC~6266 (M62),
containing, respectively, seven, six, and five type~II Cepheids within the
tidal radii of these clusters.
 
We cross-matched our list of 60 type~II Cepheids with the updated
version\footnote{\it http://www.astro.utoronto.ca/\~{}cclement/read.html} of
the catalog of variable stars in the Galactic globular clusters (Clement
\etal 2001) and we successfully identified 36 objects. The other 24 type~II
Cepheids are candidates for new cluster members. This group includes three
Cepheids in Terzan~6 (the first variable stars known in this cluster),
three additional Cepheids in Terzan~1, and the first known Cepheids in the
globular clusters NGC~6355, NGC~6540, NGC~6749, Terzan~9, and Djorg~2.

Anomalous Cepheids are extremely rare in globular clusters. The Clement
\etal (2001) catalog contains only five such objects, including BL~Bootis
(V19 in NGC~5466) -- the prototype of a class of anomalous Cepheids. Our
collection includes three anomalous Cepheids (all of them are
first-overtone pulsators) positionally coincident with globular
clusters. One of these stars (OGLE-GAL-ACEP-082 = NGC~6304 V4) has already
been classified in the OCVS as an RR~Lyr star (OGLE-BLG-RRLYR-00225). In
this investigation, we change this classification following the suggestion
of De~Lee \etal (2006), who noticed a characteristic light curve shape of
this star. The two remaining anomalous Cepheids located within tidal radii
of globular clusters are new discoveries. OGLE-GAL-ACEP-100 seems to be a
member of NGC~6541, considering its brightness compared to the brightness
of RR~Lyr stars in the same cluster. In turn, OGLE-GAL-ACEP-105 is too
bright to be a member of NGC~6638 -- we assume that this is a field
anomalous Cepheid located by chance in the area outlined by the cluster
radius.

On the other hand, our data clearly show that the variable star V12 in
globular cluster NGC~6333 (M9) is a type~II Cepheid (OGLE-BLG-T2CEP-1158),
contrary to the classification of this object as an anomalous Cepheid
suggested by Arellano Ferro \etal (2013). The light curve shape of V12, its
Fourier parameters, and its brightness in relation to the brightness of
RR~Lyr stars in this cluster undoubtedly indicate that this object belongs
to the BL~Her class.

Finally, our collection contains four stars classified as classical
Cepheids in the regions outlined by the tidal radii of globular clusters:
OGLE-BLG-CEP-034 in NGC~6355, OGLE-BLG-CEP-068 in Pal~6, OGLE-BLG-CEP-098
in NGC~6569, and OGLE-GD-CEP-1244 in GLIMPSE01. The cluster membership
status of these variables will be a matter of further research.

\Section{Conclusions}
We presented a sample of 742 additional Galactic Cepheids found in the OGLE
GVS fields. Nearly 70\% objects from this sample are new discoveries. We
also supplemented photometric time series of the previously published
Cepheids and removed some misclassified objects. This is a final release of
the OGLE Collection of Galactic Cepheids.

The huge OGLE photometric database is invariably a source of important
astrophysical discoveries. In this paper we presented, among others, the
longest-period beat Cepheid, the first known Galactic double-mode 2O/3O
classical Cepheids, candidates for single-mode first-overtone type~II
Cepheids, and new eclipsing binary systems with pulsating components.

\Acknow{We are grateful to R.~Smolec for helpful suggestions which improved
  the paper. We thank A.~Schwar\-zenberg-Czer\-ny for providing software
  used in this study.

  This work has been supported by the National Science Centre, Poland,
  grant MA\-ESTRO no. 2016/22/A/ST9/00009. The OGLE project has received
  funding from the Polish National Science Centre grant MAESTRO
  2014/14/A/ST9/00121. This research has made use of the International
  Variable Star Index (VSX) database, operated at AAVSO, Cambridge,
  Massachusetts, USA.}

\end{document}